\documentclass[preprint,times]{elsarticle}
\journal{Computer Physics Communications} 
\usepackage[utf8]{inputenc}
\usepackage{braket}
\usepackage{amsmath}

\begin{document}

\title{Implementation and performance analysis of bridging Monte Carlo moves
for off-lattice single chain polymers in globular states}

\author[mainz]{Daniel Reith\corref{cor1}}
\ead{reithd@uni-mainz.de}

\author[mainz]{Peter Virnau}
%\ead{virnau@uni-mainz.de}

\cortext[cor1]{Corresponding author}
\address[mainz]{Institut für Physik, Staudingerweg 7, 55099 Mainz, Germany}

\begin{abstract}
Bridging algorithms are global Monte Carlo moves which allow for an efficient sampling
of single polymer chains.
In this manuscript we discuss the adaptation of three bridging algorithms from lattice to
continuum models, and give details on the corrections to the acceptance 
rules which are required to fulfill detailed balance. For the first time we are able to
compare the efficiency of the moves by analyzing the occurrence of knots in globular states. For a
flexible homopolymer chain of length $N=1000$, independent configurations can be generated up
to two orders of magnitude faster than with slithering snake moves. 
\end{abstract}

\begin{keyword}
off-lattice Monte Carlo simulations\sep globular polymers \sep knots
\end{keyword}
\maketitle
\section{Introduction}
The estimation of thermodynamic quantities for globular phases of single
chain polymers remains a challenging task \cite{rensburg2009, binderpaulreview08,sokal1996}
because an effective sampling is hindered by self--entanglements, knots \cite{virnau_05, grosberg_93, mansfield_94, grosberg2009, orlandini_09} and high
density.
Globular polymers display an interesting phase behavior \cite{paul_05, taylor_09} and may serve as model systems to study
aspects of biological macromolecules \cite{grosberg_crumpled, grosberg_06, wuest_09, virnau_06}.
%The pivot algorithm \cite{lal1969, macdonald1985, madras1988}, e.g.,
%which works particularly well for polymers in good solvent conditions becomes
%inadequate for higher densities because it becomes increasingly more difficult
%to suggest configurations without creating significant overlaps. 
Typically, approaches like the slithering snake algorithm \cite{kron1965, kron1967, wall1975, mandel1979} or various types of chain growth
algorithms like PERM \cite{perm} are applied to sample such configurations. 

A new class of Monte Carlo moves which takes advantage of destroying and
reconnecting bonds was first suggested independently by Mansfield  \cite{mansfield1982} and Olaj and Lantschbauer \cite{olaj1982} in 1982. 
Even though both papers focus on sampling (polydisperse) lattice melts, Ref. \cite{mansfield1982} also includes 
the so called backbite move which preserves monodispersity and can be applied to single chains as well.
Two additional bridging moves for single polymer 
chains on the lattice were subsequently introduced by Deutsch in 1997 \cite{deutsch1997} and have since been combined with schemes to sample
free energy landscapes \cite{wuest_09, paul_05} like Wang-Landau sampling \cite{wanglandau2001}.
Nowadays, a rather evolved set of bridging methods exists for off-lattice atomistic polymer melts
\cite{theodorou1995, uhlherr2000_macro, wick2000_macro, uhlherr2001, theodorou2002_book, mavrantzas2005}, and one particular move was also ported to 
single chain simulations \cite{theodorou2002_jcp}.
Apart from the obvious application of sampling static melt quantities, these moves are also applied 
to generate well-equilibrated starting configurations for subsequent
Molecular Dynamics runs \cite{auhl, baschnagel_08}. 

Our paper focuses on the efficient generation of globular off-lattice polymers. To this extent 
we adapt the three single chain bridging moves of
Mansfield and Deutsch \cite{deutsch1997} to continuum models.
Technical aspects of our implementation which further improve
efficiency are also discussed.
In the second part we test the performance of the moves
and compare it to a standard implementation of the slithering snake algorithm.
Distinguishing independent configurations in globular states is a formidable
task on its own as commonly used observables are typically unable to yield information about
the topology of the chain. To this end, we determine whether or not a globular state 
is knotted and compare the times which are required to generate ``topologically independent''
configurations.

\section{Description of the bridging moves and detailed balance}
%

%Following this definition two bonds are then said to be parallel
%neighbors if
%the starting monomer and the ending monomers of the bonds are neighbors. It is
%also possible that the starting monomer of bond I is a neighbor of the
%ending monomer of  bond II and the ending monomer of bond I of the
%starting monomer of the bond II. Bond I and II are then said to be anti parallel
%neighbors. Depending if the bonds are parallel or anti parallel neighbors,
%internal rebridging of type I or type II is possible. Sometimes two bonds are
%parallel and antiparallel (figure \ref{fig:parallelantiparallel}\footnote{For
%simplicity the three dimensional configurations are projected into an arbitrary
%plane in all illustrations.}). Both types of
%internal rebridging steps can be done in this case.     
%
%After defining neighboring monomers and bonds off-lattice, the second point to
%take care of is the fact, that the number of neighboring monomers or neighboring
%bonds is not bounded from above. As a consequence the proposal probability is
%not symmetric. If the system is in state $i$ the probability suggesting state
%$j$ will be different from suggesting state $j$ out of state $i$. The acceptance
%probabilities for the different moves have to be corrected in that way, that
%detailed balance is fulfilled despite of the asymmetric proposal probabilities.
%

\subsection{Backbite move}
\begin{figure}
\includegraphics[width=1.0\textwidth]{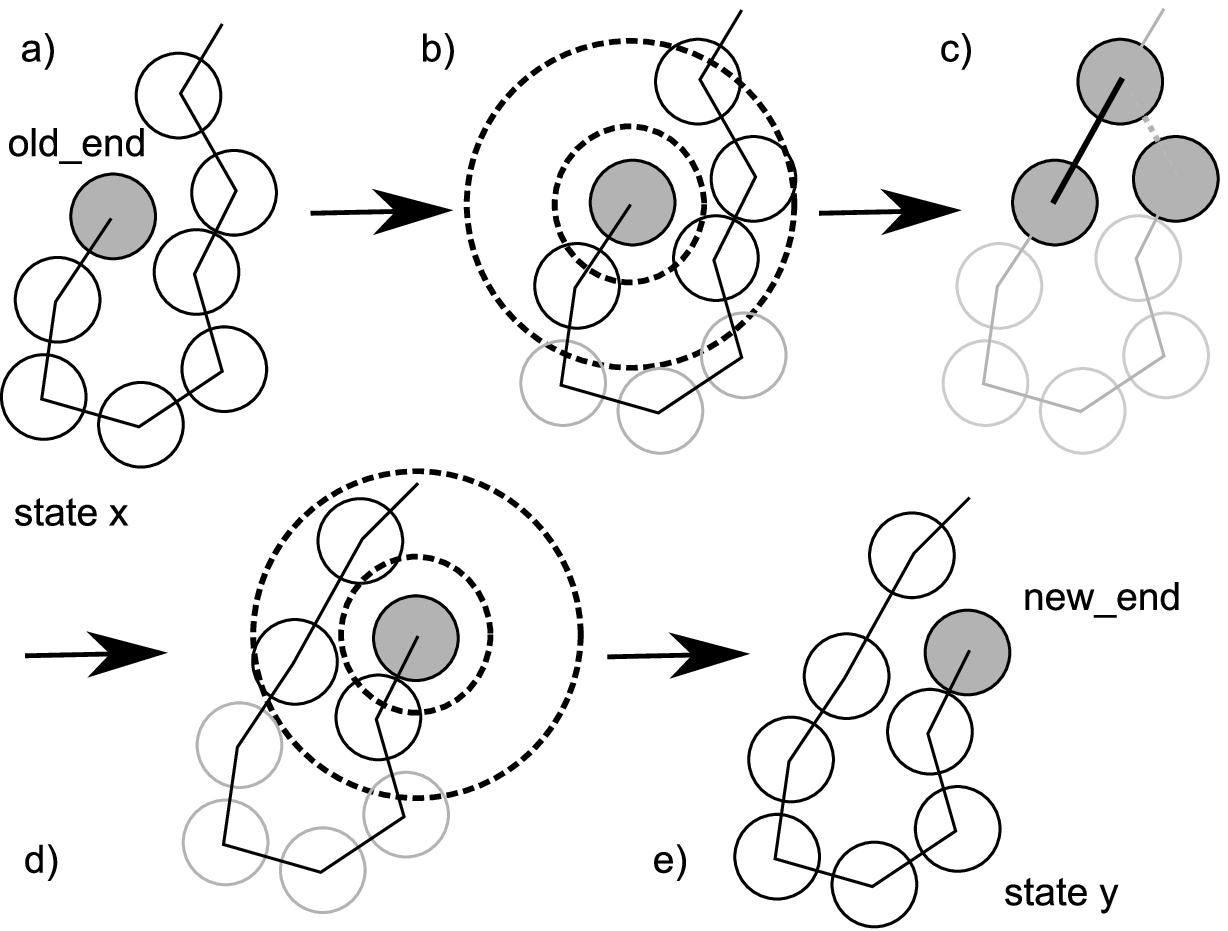}
\caption{Off-lattice version of the backbite move \cite{mansfield1982}.\newline
a) An end monomer is chosen with equal probability. \newline
b) Neighbors (within distance  $d_{min} < d < d_{max}$) of this monomer are identified and counted. 
One of these neighbors is selected at random. \newline 
c) A new bond is created between the neighbor and the old end monomer. The bond between the chosen neighbor and its
successor (along the direction from the old end to the neighbor) is cut. \newline
d) The number of neighbors of the new end needs to be counted to fulfill detailed balance. \newline
e) The chosen neighbor becomes the new end and the new configuration is accepted with a modified Metropolis criterion. } \label{fig:begend}
\end{figure}

On a lattice, the number of neighbors of a particular monomer is always well-defined.
In the continuum, we call two monomers neighbors if the distance $d$ between them
is within a certain range $d_{min} < d < d_{max}$.
The basic idea of the backbite move is illustrated in figure~\ref{fig:begend}. 

First, we select an end monomer with equal probability. Neighbors of this monomer
are identified and counted ($n_{old\_end}$), and one of these neighbors is chosen at random. 
(Fig.~\ref{fig:begend}b). A new bond is created between the neighbor and the
old end monomer. At the same time, the bond between the chosen neighbor and its
successor (along the direction from the old end to the neighbor) is severed, 
turning the neighbor into the new terminus (Fig.~\ref{fig:begend}c). Note that the selection probability
for the reverse Monte Carlo move $a_{yx}$ (from $y$ to $x$) may be different from $a_{xy}$. 
Therefore, we also need to count the number of neighbors of the new end 
$n_{new\_end}$ (Fig.~\ref{fig:begend}~d):
\begin{equation}
a_{xy} = \frac{1}{2} \cdot \frac{1}{n_{old\_end}} \text{ and } a_{yx} =
\frac{1}{2} \cdot \frac{1}{n_{new\_end}},
\end{equation}
Moves are accepted with a modified Metropolis criterion:
\begin{equation}
r <\min{\left[1,
\frac{n_{old\_end}}{n_{new\_end}} \exp{\left(-\beta \Delta V\right)
}\right]},
\end{equation}
where $r \in (0,1)$ is a random number between 0 and 1. The prefactor
$\frac{n_{old\_end}}{n_{new\_end}}$ ensures that the asymmetric proposal probability is
corrected. $\Delta V$ denotes the energy difference between the new state $y$
and the old state $x$.

The backbite move was first suggested by Mansfield for a lattice model \cite{mansfield1982}. 
Note that for a lattice model, the number of neighbors need not be counted necessarily. One can
simply choose a potential site and reject the move if there is no monomer present. 
Successful implementations in the continuum were reported recently in Refs.~\cite{taylor_09, schnabel_09}.

\subsection{Internal long range move of type II}
\begin{figure}
\includegraphics[width=1.0\textwidth]{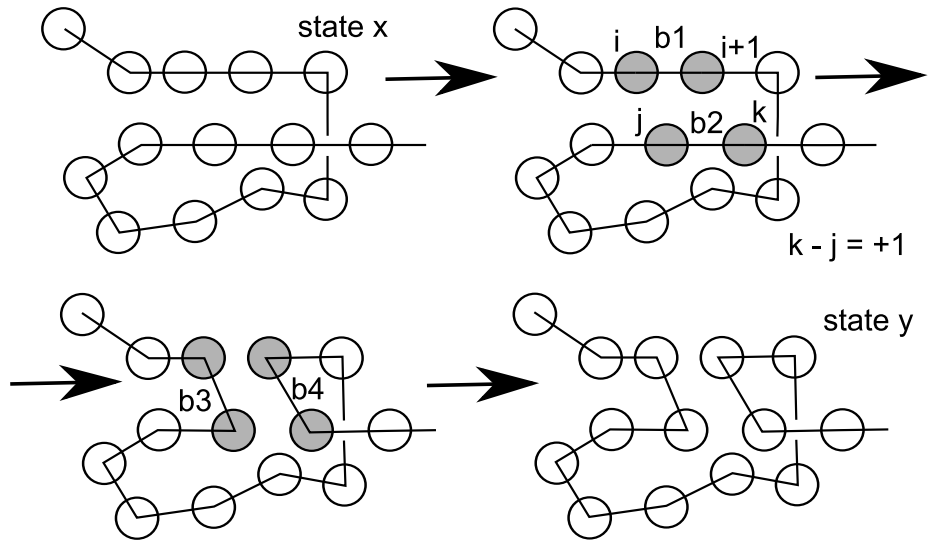}
\caption{Internal long range move of type II (parallel bonds: k-j=+1)
For details see main text.}\label{fig:internal1}
\end{figure}

\noindent
In Ref.~\cite{deutsch1997} Deutsch proposes two additional moves for single chains in 
globular states, which were originally used to search for ground-states in the
HP-model \cite{hpmodel2}. Both Monte Carlo moves cut and ``rewire'' the polymer internally, and
have to our knowledge not been ported to a continuum model so far.
First, we recall Deutsch's definition of parallel and anti-parallel bonds: 
Two opposing strands along the chain made up of monomers $i,i+1$ and $j,k$ are named parallel
if $k-j=+1$ and anti-parallel if $k-j=-1$ (see figs.~\ref{fig:internal1},~\ref{fig:internal2}). In the continuum monomers
which are about to be rewired need to be located within $d_{min} < d < d_{max}$. In this notation new bonds are always created between $i$ and $j$ and $i+1$ and $k$, which need to be neighbors.
When bonds are anti-parallel (and not parallel as suggested
in the original publication), we attempt a move of type I.
Likewise, if bonds are parallel we try a type II move. 
For pedagogical reasons we will first discuss moves of type II (Fig.~\ref{fig:internal1}) because they are easier to implement.

First we select one bond $b_1$ made up of monomers $(i,i+1)$ at random (with probability $1/(N-1)$) and choose
one of $n_{b1}$ neighboring parallel bonds $b_2$.
If no such bonds
exist, the move is rejected. Bonds $b_1$ between $i$ and $i+1$ and between $b_2$ between 
$j$ and $k$ are cut and reconnected to $b_3$ between $i$ and $j$ and $b_4$ between $i+1$ and $k$.
This leads to following selection probabilities from state $x$ to $y$:
\begin{equation}
a_{xy} = \frac{1}{N-1} \left( \frac{1}{n_{b_1}} + \frac{1}{n_{b_2}}\right).
\end{equation}
Note that the same configuration could have been chosen by first selecting $b_{2}$ instead of $b_1$, hence we
get the additional summand.
For the reverse move, we obtain
\begin{equation}
a_{yx} = \frac{1}{N-1} \left( \frac{1}{n_{b_3}} +
\frac{1}{n_{b_4}}\right).  
\end{equation}  
A move is accepted if
\begin{equation}
r <\min{\left[1,  \frac{n_{b_1} n_{b_2} ( n_{b_3} + n_{b_4})}{n_{b_3} n_{b_4}
(n_{b_1} + n_{b_2})}\exp{\left(-\beta\Delta V\right)
}\right]}.
\end{equation}
As before, $\Delta V$ denotes the energy difference between the new state $y$
and the old state $x$.
Note that the selection probabilities are symmetric in the original publication of Deutsch \cite{deutsch1997}.
One simply checks for neighboring bonds on the lattice and rejects the move if no adequate bond is present.

\subsection{Internal long range move of type I}

\begin{figure}
\includegraphics[width=1.0\textwidth]{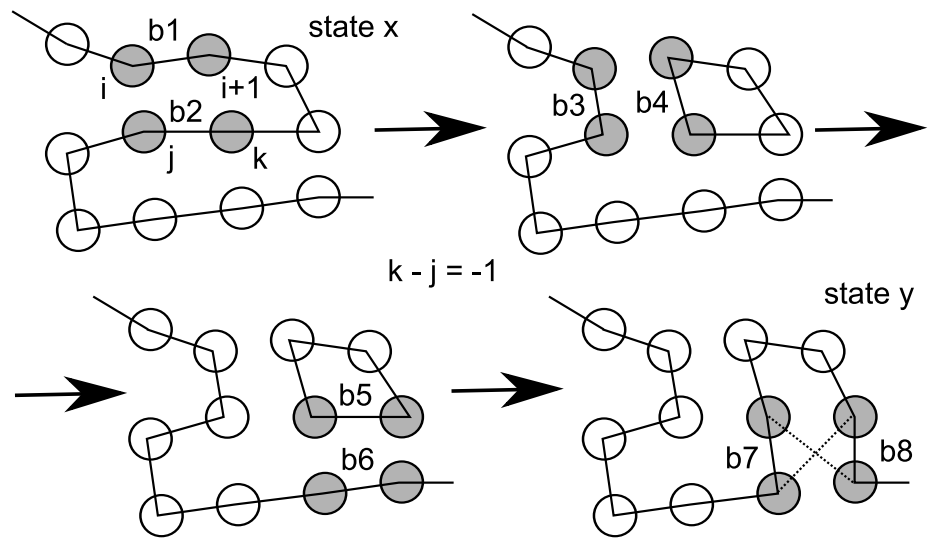}
\caption{Internal long range move of type I (anti-parallel bonds: k-j=-1):
The chain is split into a linear and a circular part and reconnected.
For details see main text.}\label{fig:internal2}
\end{figure}

Figure \ref{fig:internal2} provides details on the long range move of type I.
First we select one bond $(i,i+1)$ at random (with probability $1/(N-1)$) and choose
one of the neighboring anti-parallel bonds $b_2$ (with probability $1/n_{b1}$).
Bonds $b_1$ between $i$ and $i+1$ and between $b_2$ between $j$ and $k$ are cut and
reconnected to $b_3$ between $i$ and $j$ and $b_4$ between $i+1$ and $k$ such that the chain
is split into a linear and a circular part. From the circular part which consists of $n_z$ bonds we choose
a random bond ($b_5$ in fig.~\ref{fig:internal2} with probability $1/N_z$) 
and check if neighboring bonds belong to the linear part. One of these $n_{b5}'$ bonds is chosen at random ($b_6$)
and the chain is ``rewired'' to form the anti-parallel bonds $b_{7}$ and $b_{8}$. Sometimes it is possible to reconnect bonds $b_5$ and $b_6$ in different ways (as indicated by dotted lines in fig.~\ref{fig:internal1}). In this case each potential connection contributes to $n^\prime_{b_5}$. If there is no such bond $b_6$, 
the move is rejected. Note that in all steps, the number of neighboring bonds need to be determined. 
This leads to following selection probabilities from state $x$ to $y$:
\begin{equation}
a_{xy} = \frac{1}{N-1} \cdot \frac{1}{n_z} \cdot \frac{1}{n^\prime_{b_5}}
\left( \frac{1}{n_{b_2}} + \frac{1}{n_{b_1}}\right)
\end{equation}
Note that the same configuration could have been chosen by first selecting $b_{2}$ instead of $b_1$, hence we
obtain the additional summand.
For the reverse move, we get
\begin{equation}
a_{xy} = \frac{1}{N-1} \cdot \frac{1}{n_z} \cdot \frac{1}{n^\prime_{b_4}}
\left( \frac{1}{n_{b_8}} + \frac{1}{n_{b_7}}\right)
\end{equation}
This leads to the following Metropolis criterion:
\begin{equation}
r<\min{\left[1,
\frac{n_{b_1} n_{b_2} n^\prime_{b_5} (n_{b_7} + n_{b_8})}{n_{b_7} n_{b_8}
n^\prime_{b_4}(n_{b_1} + n_{b_2})} \exp{\left(-\beta \Delta V\right)
}\right]}.
\end{equation}
As before, $\Delta V$ denotes the energy difference between the new state $y$
and the old state $x$.
\subsection{Implementation}
In our program monomers of the chain are stored in sequence in a double-linked list. 
This allows for an efficient detection of parallel and anti-parallel bonds 
for internal moves of type II and I.
Even more importantly, we set up neighbor tables in which the neighbors of each monomer are stored. 
Unfortunately, this neighbor table needs to be updated each time local moves, which are essential
for the ergodicity of the algorithms, are attempted. In order to minimize these
efforts, we always perform several local displacements for all particles before
we return to a sequence of bridging moves. 

As indicated in previous sections, the internal move of type I is by far
the most complicated to implement because we need to take care of the chain separation in
both linear and circular parts. The easiest move is implement is the backbite move
followed by the internal bridging move of type II.

\section{Model and performance analysis}
In the following we compare the performance of the moveset with an implementation of
the slithering-snake algorithm \cite{wall1975}: We choose with probability 0.5 one end of the chain
and attempt to attach the end monomer to a random position at the opposite terminus without
changing the bond length. Our model system consists of a simple flexible
Lennard-Jones + FENE homopolymer \cite{kremer_1990, virnau_05, virnau_jcpco2} with cut and shifted Lennard-Jones
interactions between all monomers 
\begin{equation}
V_{LJ} = 
\begin{cases} 
4 \epsilon \left[ \left(\frac{\sigma}{r}\right)^{12} - \left(\frac{\sigma}{r}\right)^6 + \frac{127}{16384}\right] 
& \text{, if } r < 2 \sqrt[6]{2} \\ 
0 & \text{, otherwise} \label{eqn:lj}
\end{cases}
\end{equation}
and FENE interactions between adjacent beads:
\begin{equation}
V_{FENE} = - 33.75 \epsilon \ln{\left[ 1 - \left( \frac{r}{1.5
\sigma}\right)^2\right]. }\label{eqn:fene}
\end{equation}
Note that for a homopolymer as described in Eqs.~\ref{eqn:lj} and \ref{eqn:fene}, the energy difference 
in the Metropolis criterion of the bridging moves only consists of
changes in the bond energy as the position of the particles are only altered by local moves.
All simulations took place at $T=1.66~\epsilon/k_{B}$, which corresponds to $0.5T_{\Theta}$ (for
a chain of infinite length \cite{bindervirnau_review}.) Therefore, for $N=200$ to $N=1000$ only 
globular states are observed.

In an attempt to gauge the efficiency of the three bridging moves, we 
determined whether or not a particular globule is unknotted by calculating 
the Alexander polynomial \cite{virnau_knotdetection}. Note that mathematically knots
are only well-defined in closed curves. However, if we connect the ends of an open chain 
in a well-defined manner we can still obtain corresponding information about the 
``knottedness'' of the globule. 
Ref.~\cite{virnau_knotdetection} provides details on
this topic as well as on our implementation of the Alexander polynomial. The particular 
closure used in this work is described in Ref.~\cite{virnau_06}.
To quantify the number of independent configurations generated by the bridging 
algorithms we measured the correlation times between unknotted configurations \cite{landaubinder}:
\begin{equation}
A(k) = \frac{\braket{U_i U_{i+k}} - \braket{U_i}^2}{\braket{U_i^2} - \braket{U_i}^2}
\end{equation}
where 
\begin{equation}
U_i = \begin{cases} 0, & \text{ if the $i$th configuration contains a knot} \\ 
1, & \text{if the $i$th configuration is unknotted.}\end{cases}
\end{equation}
An ``independent configuration'' is defined by the decrease of the auto-correlation function to $e^{-1}$.
Note that our approach is particularly well-suited for the task of comparing bridging moves as it 
focuses on changes in topology and self-entanglements.
In our current implementation neither bridging moves
nor slithering snake moves are ergodic because they do not alter bond lengths. Therefore,
it becomes necessary to combine the algorithms with local Monte Carlo moves: In our case, 
each simulation run spent 26 $\%$ of the time with local displacements.

\begin{figure}
\includegraphics[width=1.\textwidth]{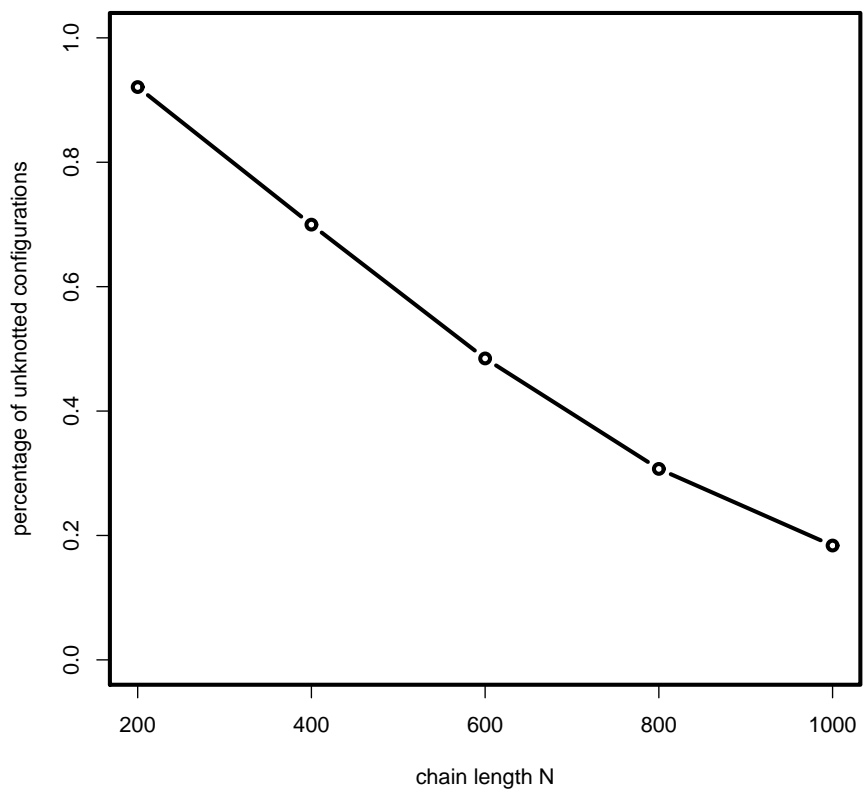}
\caption{Percentage of unknotted configurations for single bead-spring polymer chains of size N=200 to N=1000 in the globular phase ($T=1.66~\epsilon/k_B$) }\label{fig:unknottingpercentage}
\end{figure}

\begin{figure}
\centering
\includegraphics[width=0.75\textwidth]{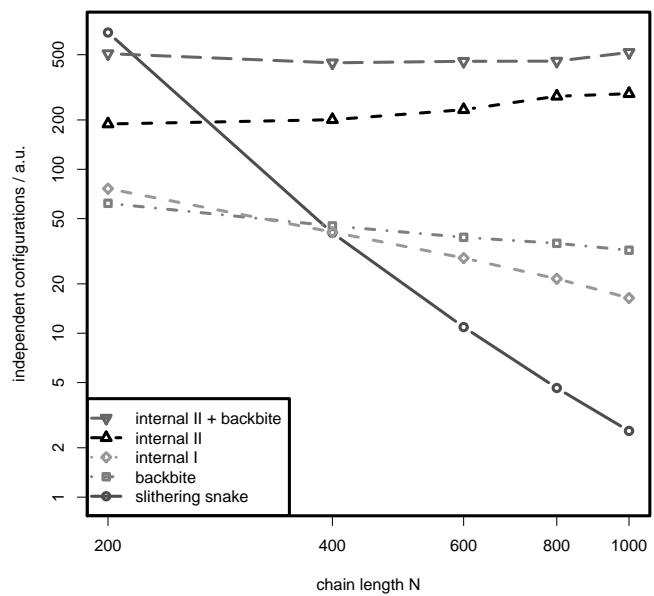}
\includegraphics[width=0.75\textwidth]{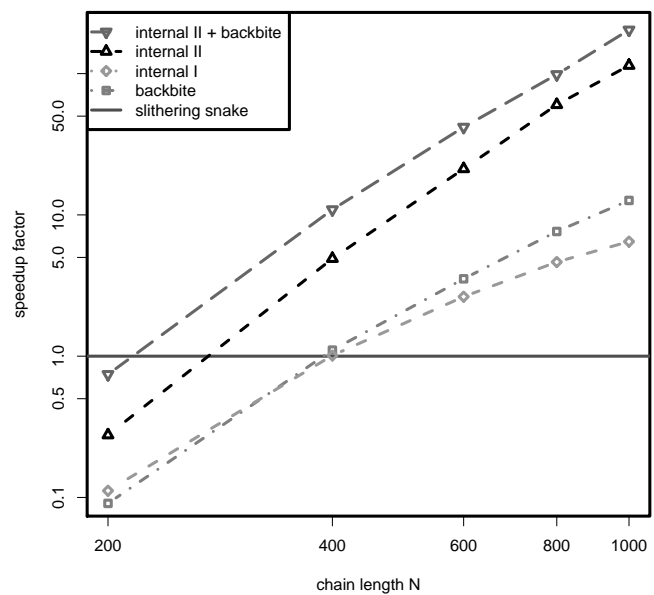}
\caption{Performance analysis for single bead-spring polymer chains of size N=200 to N=1000 in the globular phase ($T=1.66~\epsilon/k_B$)  \newline
a. Number of ``independent configurations'' per minute as defined in the main text. \newline
b. Speed-up factor of bridging moves in comparison with slithering snake moves.
For this particular model all rebridging moves become more efficient than slithering snake 
moves for chains larger than 400 monomers. For clarity we have used logarithmic scale on both axis.}\label{fig:performance1}
\end{figure}

Fig.~\ref{fig:unknottingpercentage} shows the percentage of unknotted configurations for chain lengths ranging from $N=200$ up to $N=1000$: For $N=200$ roughly 90 \% of all configurations are unknotted, for $N=1000$ only 20 \% are unknotted \cite{virnau_05}. Figure~\ref{fig:performance1}a shows the number of generated configurations per minute divided by our estimate
for the geometric correlation time as a function of chain length. All configurations were created on a single 
core of a Core2Quad Q6700 processor running at 3.33 Ghz. Figure~\ref{fig:performance1}b shows the same data normalized
by the results for the slithering snake moves. Note that the time-limiting step in this analysis is not the generation of
globular states but the subsequent knot analysis which took place on a supercomputer.
For chains larger than $N=400$ all bridging moves become more efficient than slithering snake moves.
The internal bridging move of type II is the most efficient followed by the backbite move and the internal move of type I.
If we consider that the latter is by far the most difficult to code among the three, we do not recommend its implementation.
Interestingly, the time to generate an ``independent configuration'' appears to be almost independent of chain length for 
type II and backbite moves, whereas it increases rapidly for slithering snake moves as expected.
Finally, it is worth noting that combinations of moves may even yield better results. 
A combination of internal II and backbite moves is about two orders of magnitude faster for $N=1000$ than slithering snake moves.

\section{Conclusion}
The aim of this paper is threefold. First, we have implemented three efficient lattice bridging moves for single flexible 
globular homopolymer chains in a continuum model, and discussed difficulties arising from the implementation. In an attempt
to measure performance we determined correlation times between unknotted globular states. To our knowledge this is the first time
in which topology is considered to gauge the efficiency of global Monte Carlo moves. From this analysis, we conclude that
all bridging moves become more efficient than slithering snake moves for chains larger than $N=400$ monomers. Among the three moves,
the internal bridging move of type II is the most successful. The most complicated move (internal I) performs worst and we do not recommend its
implementation. For $N=1000$ a combination of backbiting and internal bridging moves of type II is two orders of magnitude faster
than our implementation of the slithering snake move.

The exact speed--up factors should, however, be taken with a grain of salt and rather serve as a guideline.
Long-range correlations, which are only relaxed by local moves, were e.g. not considered, and results also depend a lot on the model under 
consideration. If chain stiffness is included \cite{cifra_08}, efficiency 
will drop, and results will also deteriorate if the implementation does 
not include look-up tables. Nonetheless, we believe to have demonstrated that bridging algorithms are indeed the method of choice for the simulation of 
long globular polymers in the continuum.
 
\section*{Acknowledgments}
We would like to particularly thank T. Wüst whose comments helped us tremendously to implement and further improve the bridging moves.
We would also like to thank K. Binder, W. Paul and M. Taylor for fruitful discussions on the topic.
This work received financial support from the Deutsche Forschungsgemeinschaft (SFB 625-A17) and the
MWFZ Mainz. Computing time on JUMP and JUROPA supercomputers at the NIC Jülich and on the linux cluster at the ZDV
Mainz are gratefully acknowledged, too.

%\bibliography{literatur}

%\begin{figure}
%\includegraphics[width=1.0\textwidth]{./grafiken/parallelantiparallel.eps}
%\caption{
%It is possible to cut and reconnect neighboring bonds of a single chain 
%in two ways if the distance between the monomers involved is appropriate.
%Creating new bonds between the two starting monomers (S) and the two end monomers (E) 
%of the old bonds is referred to as type I, while creating two new S-E bonds is referred to as type
%II.}\label{fig:parallelantiparallel} 
%\end{figure}
\clearpage

\newpage
\clearpage
\thispagestyle{empty} 

\end{document}